\documentclass{jfm}

\usepackage{graphicx}
\usepackage{dcolumn}
\usepackage{bm}
\usepackage{amssymb}
\usepackage{CJK}
\usepackage{amsmath,mathtools}
\usepackage{color}
\usepackage{hyperref}
\usepackage{psfrag}
\usepackage{stmaryrd}
\usepackage{amssymb}
\usepackage{wasysym}
\usepackage[utf8]{inputenc}
\usepackage{tabularx}
\usepackage{multirow}
\usepackage{natbib}

\newcommand{\ubd}{{\bf u}_\text{2D}}
\newcommand{\hutd}{\hat{{\bf u}}_\text{3D}}

\shorttitle{Three-dimensionality in rapidly rotating flows}
\shortauthor{K. Seshasayanan and B. Gallet}

\title{Onset of three-dimensionality in rapidly rotating turbulent flows}

\author{Kannabiran Seshasayanan and Basile Gallet}

\affiliation{Service de Physique de l'\'Etat Condens\'e, CEA,
CNRS UMR 3680, Universit\'e Paris-Saclay, CEA Saclay, 91191 Gif-sur-Yvette,
France}

\date{\today}

\begin{document}

\maketitle


\begin{abstract}
Turbulent flows driven by a vertically invariant body force were proven to become exactly two-dimensional above a critical rotation rate, using upper bound theory. This transition in dimensionality of a turbulent flow has key consequences for the energy dissipation rate. However, its location in parameter space is not provided by the bounding procedure. To determine this precise threshold between exactly 2D and partially 3D flows, we perform a linear stability analysis over a fully turbulent 2D base state. This requires integrating numerically a quasi-2D set of equations over thousands of turnover times, to accurately average the growth rate of the 3D perturbations over the statistics of the turbulent 2D base flow. We leverage the capabilities of modern GPUs to achieve this task, which allows us to investigate the parameter space up to $Re=10^5$. At Reynolds numbers typical of 3D DNS and laboratory experiments, $Re\in[10^2, 5\, 10^3]$, the turbulent 2D flow becomes unstable to 3D motion through a centrifugal-type instability. However, at even higher Reynolds number another instability takes over. A candidate mechanism for the latter instability is the parametric excitation of inertial waves by the modulated 2D flow, a phenomenon that we illustrate with an oscillatory 2D Kolmogorov flow.
\end{abstract} 

\section{Introduction}

Global rotation tends to make turbulent flows two-dimensional and invariant along the rotation axis. This result is often referred to as the Taylor-Proudman theorem, which states that slowly evolving large-scale rapidly rotating flows organise into Taylor columns, invariant along the rotation axis (the vertical axis thereafter). However, in most numerical and experimental studies of rapidly rotating turbulence these large scale columns coexist with erratic 3D small-scale motion \citep{Bartello1994,Yeung1998,smith1999transfer,Chen2005,Morize2006,Staplehurst2008,Thiele2009,moisy2011decay,
gallet2014scale,Yarom2013,yokoyama2017hysteretic,seshasayanan2018condensates}. Although the latter contain typically less kinetic energy than the large-scale columns, they need to be precisely characterised, because they are responsible for the efficient 3D dissipation of kinetic energy \citep{CampagnePRE}. Indeed, at the theoretical level the 3D inertial waves have been shown to induce a direct wave turbulent cascade leading to `anomalous' kinetic energy dissipation, i.e., a dissipated power independent of molecular viscosity \citep{Galtier,Bellet}. On the one hand, WT provides the right framework to address the decay of rotating turbulence initialised with inertial waves only. On the other hand, forced rotating turbulence in a finite domain generically exhibits intense Taylor columns when global rotation is sufficiently fast, which challenges a description in terms of inertial waves only (even when energy is input into wave modes only, see \citet{Brunet2020, Reun2020}).

We focus on such forced rotating turbulence in statistically steady state. Restricting attention to a steady body-force at scale $L$, the goal  is to charaterise the turbulent flows arising at large Reynolds number $Re= U L / \nu$, where $U$ is the root-mean-square (rms) flow velocity and $\nu$ is the kinematic viscosity, and low Rossby number $Ro=U/2\Omega L$, where $\Omega$ denotes the global rotation rate. Two situations emerge depending on the spatial structure of the body-force: 
\begin{itemize}
\item When the body-force is three-dimensional and directly drives some vertically dependent flow structures, it was shown by \cite{alexakis2015rotating} that the statistically steady state never corresponds to a rapidly rotating turbulent flow with both $Re \gg 1$ and $Ro \ll 1$. The reason is that the rms velocity $U$ is an emergent quantity that cannot be specified at the outset of a numerical simulation. In (non-turbulent) flows characterised by $Re \sim 1$, the flow can achieve arbitrarily low $Ro$ provided the global rotation is fast enough. However, for turbulent flows with $Re \gg 1$, the rms velocity saturates at a value $U \sim \Omega L$, such that the Rossby number approaches unity.
\item By contrast, it is only when the body-force is invariant along the rotation axis that one can reach simultaneously $Re \gg 1$ and $Ro \ll 1$, a regime that we refer to as rapidly rotating turbulence. 
\end{itemize}
For the latter situation, one can prove that the flow becomes exactly two-dimensional when the Rossby number is reduced at fixed (but arbitrarily large) Reynolds number and fixed aspect ratio of the fluid domain, using rigorous upper bound theory.  This has important consequences for the energy dissipation rate of rapidly rotating turbulent flows: the anomalous dissipation associated with the forward energy cascade of 3D turbulence is replaced by the laminar-like viscous dissipation of the domain scale velocity structures that arise from the 2D inverse energy cascade \citep{alexakis2006energy,vanBokhoven2009,CampagnePOF,CampagneJFM,Deusebio2014,Buzzicotti2018,Pestana2019,van2019critical}. Mapping the energy dissipation rate of rotating turbulence in parameter space requires to determine the transition between exactly 2D and partially 3D rapidly rotating flows. Indeed, the mathematical results in \cite{gallet2015exact} only provide a lower bound on the critical Rossby number $Ro_c(Re)$ below which the flow becomes purely 2D. 
The bound typically scales as $Re^{-6}$ with logarithmic corrections, i.e., we are only able to prove that the flow becomes exactly 2D for extremely low Rossby numbers. However, one should keep in mind that this $Re^{-6}$ scaling behaviour is only a limitation of the bounding method, not a property of the \textit{true} threshold $Ro_c(Re)$: we establish in the following that $Ro_c(Re)$ is in fact much greater than $Re^{-6}$, so that exact two-dimensionalization takes place over a signification region of parameter space.

The very existence of a clear-cut transition between exactly 2D and partially 3D flows suggests an alternate approach to the study of rapidly rotating turbulence: instead of running costly 3D Direct Numerical Simulations (DNS) at large Reynolds number and ever lower Rossby number, in the present study we start from high $Re$ and very low $Ro$, where we know the flow is 2D, and investigate the appearance of three-dimensionality as we increase $Ro$. The goal is to determine the boundary $Ro_c(Re)$ in parameter space between exactly 2D flows and partially 3D ones though linear stability analysis. The challenge is that the base state of this linear stability analysis is a body-forced turbulent 2D flow. Because of the inverse energy cascade, such flows typically achieve a statistically steady state after a long transient, the duration of which scales with the viscous time scale $L^2/\nu$. For a given Reynolds number $Re\gg 1$, one typically needs to integrate the 2D equations for a number $Re$ of large-scale eddy turnover times. Once the statistically steady 2D base state is reached, thousands of additional turnover times are needed to correctly sample the growth rate of the infinitesimal 3D perturbations and conclude on the stability of the flow. We leverage the capabilities of modern GPUs to address this challenge: the linear stability problem for the 3D perturbations is effectively 2D and fits on the memory of a single GPU, which outperforms CPUs for the rapid computation of Fast Fourier Transforms. This enables us to investigate the 2D to 3D transition up to $Re=10^5$, one to two orders of magnitude above the typical values reported in experimental and fully 3D numerical studies.

\section{Theoretical setup}

\begin{figure}
\begin{center}
\centering{\includegraphics[scale=0.42]{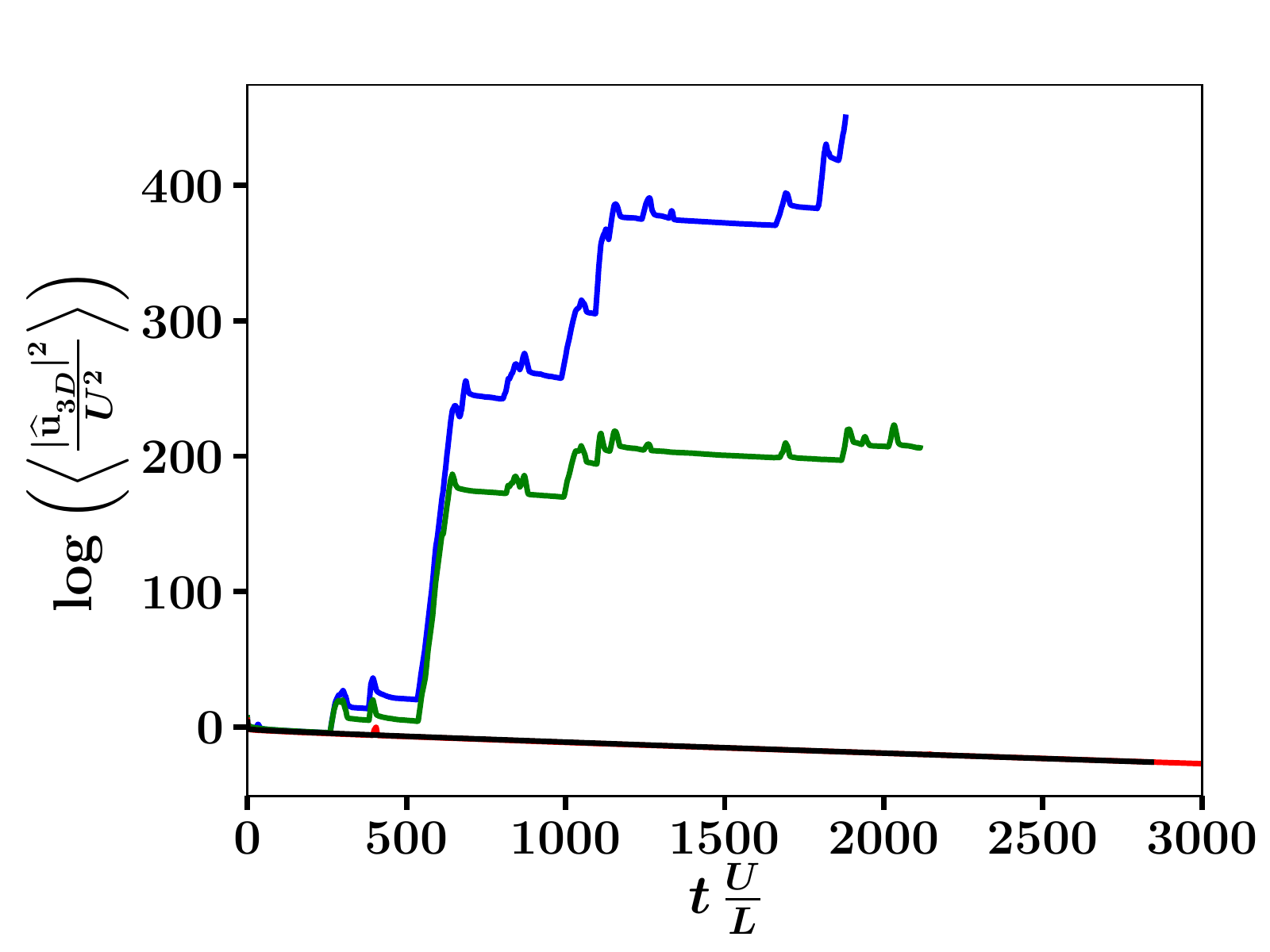}}
\end{center}
\caption{Time series of the kinetic energy of the 3D perturbation, for fixed Reynolds number $Re=9.3 \times 10^4$ and aspect ratio $qL=6 \pi$. The Rossby number increases from bottom to top, $Ro = 3.7 \times 10^{-3}$ (black), $\, 4.9 \times 10^{-3}$ (red), $\, 7.4 \times 10^{-3}$ (green) and $9.8 \times 10^{-3}$ (blue) respectively. The first two time series correspond to stable situations, while the 3D perturbation grows exponentially for the two highest values of $Ro$.}
\label{fig:timeseries}
\end{figure}

We consider the flow of a Newtonian fluid inside a parallelepipedic domain $(x,y,z) \in [0,L]\times[0,L]\times[0,H]$, in a frame rotating at a rate $\Omega>0$ around the vertical $z$ axis. We focus on periodic boundary conditions in all three directions, although the results apply equally to a fluid layer of height $H/2$ with stress-free boundary conditions at top and bottom. The flow is driven by a vertically invariant body force ${\bf f}=f_0 \sin(8\pi y/L){\bf e}_x$, where ${\bf e}_x$ denotes the unit vector along $x$, and the velocity field ${\bf u}(x,y,z,t)$ satisfies the rotating Navier-Stokes equation:
\begin{equation}
\partial_t \textbf{u} +(\textbf{u} \cdot \bnabla) \textbf{u} + 2 \Omega {\bf e}_z \times \textbf{u} =-\bnabla p + \nu \Delta \textbf{u} + {\bf f} \, ,
\label{NS}
\end{equation}
together with the incompressibility constraint $\bnabla \cdot {\bf u}=0$. This equation admits some vertically invariant 2D solutions, ${\bf u}_{2D}(x,y,t)$, that satisfy the 2D Navier-Stokes equation:
\begin{equation}
\partial_t \ubd +(\ubd \cdot \bnabla) \ubd =-\bnabla P + \nu \Delta \ubd + {\bf f} \, ,
\label{NS2D}
\end{equation}
where the Coriolis force is absorbed by the pressure gradient. In a non-rotating system and at large Reynolds number, such 2D solutions are unstable to 3D perturbations and quickly evolve into fully three-dimensional turbulence. However, rapid global rotation stabilises these 2D solutions with respect to 3D perturbations, even the turbulent ones: at large Reynolds number and very low Rossby number, the flow $\ubd(x,y,t)$ evolves in a complicated and chaotic fashion, but it remains invariant in the vertical direction. Such exact two-dimensionalization of the flow only holds up to a Reynolds-number-dependent and aspect-ratio-dependent critical Rossby number $Ro_c(Re,L/H)$ above which three-dimensionality spontaneously arises. To determine the precise threshold $Ro_c$, we consider the evolution of an infinitesimal 3D perturbation to the 2D turbulent base flow $\ubd(x,y,t)$. This 3D perturbation satisfies the Navier-Stokes equation \eqref{NS} linearised around $\ubd$. Because the latter is independent of $z$, the different vertical Fourier modes of the 3D perturbation decouple at linear order. Without loss of generality, we can thus restrict attention to a 3D perturbation consisting of a single vertical Fourier mode, $\hutd(x,y,t)e^{i q z}$, where $\hutd(x,y,t)$ is a complex-valued 3D vector. The full velocity field is ${\bf u}(x,y,z,t)= \ubd(x,y,t)+\hutd(x,y,t)e^{i q z}+ c.c.$, where $c.c.$ denotes the complex conjugate of the second term. Upon linearising the vorticity equation around the 2D base flow, we obtain the evolution equation for the vorticity of the 3D perturbation, $\hat{\bm \omega}_{\text{3D}} = \left( {\bnabla}_{\perp} + i q {\bf e}_z \right) \times \hutd$:
\begin{eqnarray}
\partial_t \hat{\bm \omega}_{\text{3D}} & = & \left( {\bm \nabla}_{\perp} + i q {\bf e}_z \right) \times \left[ {\bf u}_{\text{2D}} \times  \hat{\bm \omega}_{\text{3D}} +  \right.  \left. \hat{\bf u}_{\text{3D}} \times ( {\omega}_{_{2D}} {\bf e}_z) \right]  \nonumber \\
& & + 2 i q \Omega \, \hat{\bf u}_{\text{3D}} + \nu \left( {\bm \nabla}_{\perp}^2 - q^2 \right)  \hat{\bm \omega}_{\text{3D}}, \label{eq:omega_3D} 
\end{eqnarray}
where ${\bm \nabla}_{\perp}=(\partial_x, \partial_y, 0)$ and ${\omega}_{\text{2D}} = ({\bm \nabla}_{\perp} \times \ubd ) \cdot {\bf e}_z$.

\begin{figure}
\begin{center}
\centering{\includegraphics[scale=0.42]{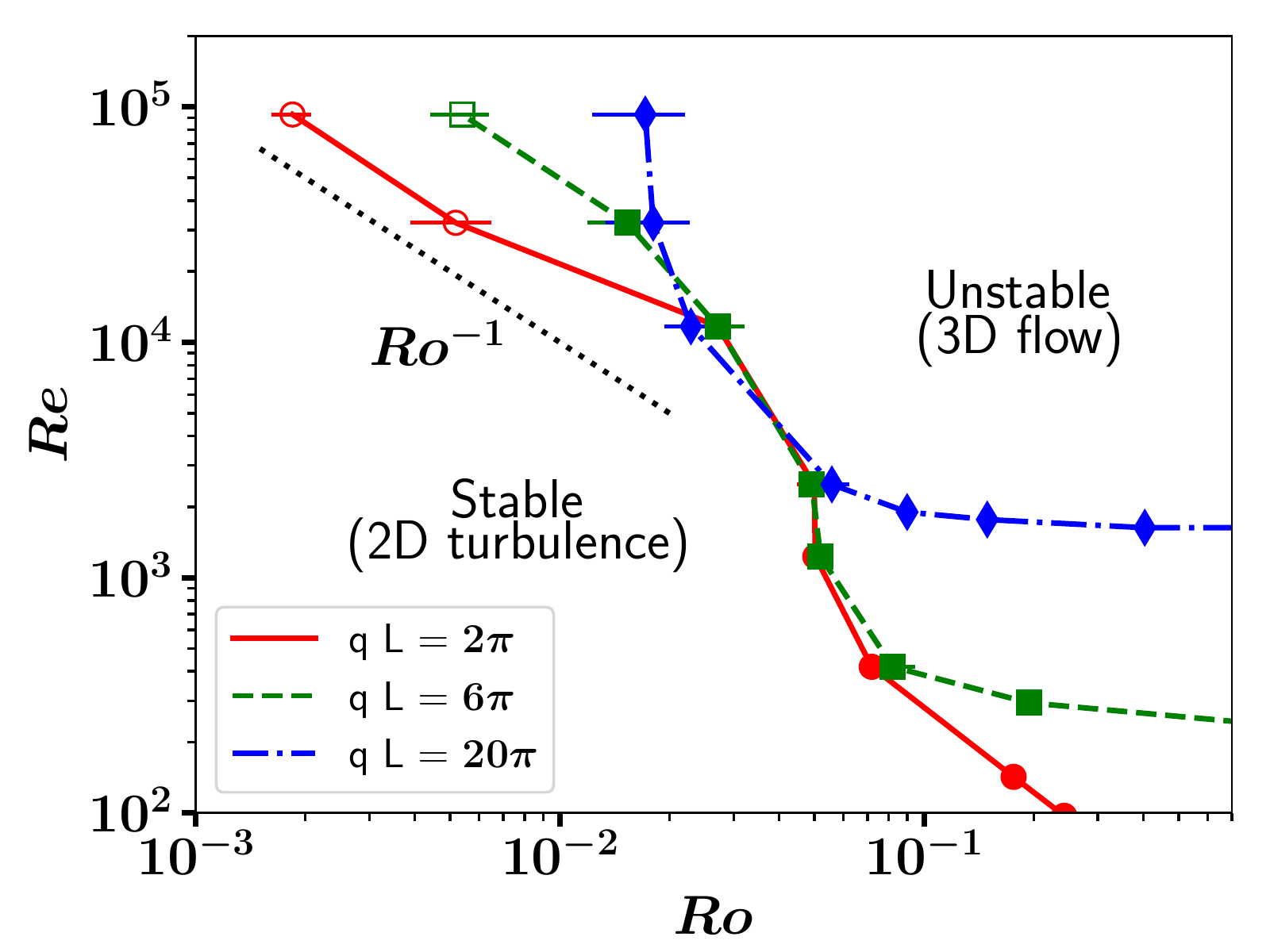}}
\end{center}
\caption{Threshold for instability to 3D perturbations in the $(Ro,Re)$ plane, for three values of the aspect ratio. Full symbols with $Ro\in [0.015,0.03]$ correspond to a centrifugal-type instability. The open symbols at lower $Ro$ depart from this behaviour, and display the characteristics of the parametric excitation of inertial waves.}
\label{fig:RocvsRe}
\end{figure}

Linear stability analysis thus boils down to an effectively two-dimensional fluid problem governed by equations \eqref{NS2D} and (\ref{eq:omega_3D}). However, the aspect ratio of the 3D domain remains a key control parameter, as it restricts the acceptable values of the vertical wavenumber $q$ entering these 2D equations. Anticipating the results presented in figure \ref{fig:RocvsRe}, we observe that $Ro_c$ is a decreasing function of the vertical wavenumber $|q|$ throughout most of the parameter space. In other words, the most unstable mode corresponds to the gravest vertical wavenumber compatible with the vertical boundary conditions, $q=2\pi/H$. Linear stability analysis for a low value of $qL$ thus yields the threshold Rossby number for a deep domain, while linear stability analysis for a large value of $qL$ yields the threshold Rossby number for a shallow domain, the dimensionless vertical wavenumber and the aspect ratio being related by $qL=2 \pi L/H$. In the following, we thus refer to the threshold Rossby number computed for large (resp. low) values of $qL$ as the onset of 3D motion in deep (resp. shallow) fluid layers.

The stability analysis consists of two steps: first, we integrate equation \eqref{NS2D} until the 2D turbulent flow reaches a statistically steady state, which constitutes the base state of the linear stability analysis. This base flow is independent of the global rotation rate $\Omega$, which does not appear in equation \eqref{NS2D}. Secondly, we introduce the 3D perturbation by solving simultaneously equations \eqref{NS2D} and \eqref{eq:omega_3D}.

We compute the root-mean-square velocity from the statistically steady 2D base flow, $U=\left< \ubd^2 \right>_{{\bf x},t}^{1/2}$, where $\left< \cdot \right>_{{\bf x},t}$ denotes an average over space and time. Even though $U$ is an emergent quantity -- as opposed to a true control parameter of equation (\ref{NS2D}) -- we build the Reynolds number $Re$ and Rossby number $Ro$ with this root-mean square velocity to facilitate comparison with other experimental and numerical setups. The relation between $Re$ and the forcing-based Grashof number associated with equation (\ref{NS2D}) is provided in Appendix \ref{appendixDNS}. The problem thus involves three dimensionless parameters: the Reynolds and Rossby numbers defined above, and the aspect ratio of the fluid domain through the dimensionless vertical wavenumber $qL$. 
A typical set of numerical runs consists in holding $Re$ and $qL$ fixed, and sweeping over the values of $Ro$ by changing the rotation rate $\Omega$. The corresponding time series of the kinetic energy of the 3D perturbation, $E_\text{3D}(t)=\left<|\hutd|^2\right>_{\bf x}$, are displayed in Fig.~\ref{fig:timeseries} for such a set of runs. For low Rossby number (large rotation rate), the 3D kinetic energy decays monotonically in time: the 2D turbulent base-flow is stable with respect to 3D perturbations. By contrast, for larger Rossby number the 3D kinetic energy grows rapidly, indicating an instability. The growth rate of the 3D kinetic energy displays a strongly intermittent behaviour, associated with the turbulent dynamics of the background 2D flow. Through long numerical integrations we infer the average growth rate, $\gamma=\left< \frac{\mathrm{d}}{\mathrm{d}t} \log E_{3D} \right>_t$, as a function of the Rossby number  $Ro$. The threshold Rossby number $Ro_c(Re,qL)$ for the emergence of three-dimensionality is obtained when $\gamma=0$. We determine $Ro_c(Re,qL)$ by repeating similar sets of numerical runs for various values of $Re$ and $qL$ and show it in Fig.~\ref{fig:RocvsRe}. The boundary clearly separates the low-$Ro$ region of parameter space, where the system has 2D-flow attractors, from the large-$Ro$ region, where the flow becomes 3D. It is worth stressing the fact that $Ro_c$ is much larger than the conservative lower bound computed in \cite{gallet2015exact}: the threshold lies in a region of parameter space accessible to DNS \citep{Godeferd1999,Mininni2009,Mininni2010} and laboratory experiments \citep{Hopfinger1982,Dickinson1983,Morize2006,Yarom,gallet2014scale}, which typically achieve $Re \leq 3000$ and $Ro \in [10^{-2},1]$. However, the integration of the quasi-2D equations on GPUs allows us to extend this boundary to more extreme values of the dimensionless parameters, all the way to $Re \simeq 10^5$ and $Ro\simeq 2. \, 10^{-3}$.

The location of the boundary depends on the aspect ratio $qL$: for a shallow fluid layer with $qL=20\pi$,  $Ro_c$ seems to asymptote to a $Re$-independent value at large $Re$, at least up to $Re=10^5$. By contrast, for deeper fluid layers (lower values of $qL$) $Ro_c$ does depend on $Re$ at large Reynolds number, with the approximate scaling behaviour $Ro_c \sim Re^{-1}$, shown as a dashed line in Fig.~\ref{fig:RocvsRe}. The instability arises over a turbulent base state, a situation less documented than standard instabilities arising over steady or periodic base flows. To gain intuition in the instability process, we thus pursue two approaches: first, we compare the present instability to known instabilities of steady vortices in rotating flows, namely the centrifugal and elliptical instabilities. Secondly, we discuss the decomposition of the low-$Ro$ 3D perturbation into inertial waves, the instabilities of which can be investigated through a perturbative expansion in Rossby number.

\begin{figure*}
\begin{center}
\includegraphics[scale=0.3]{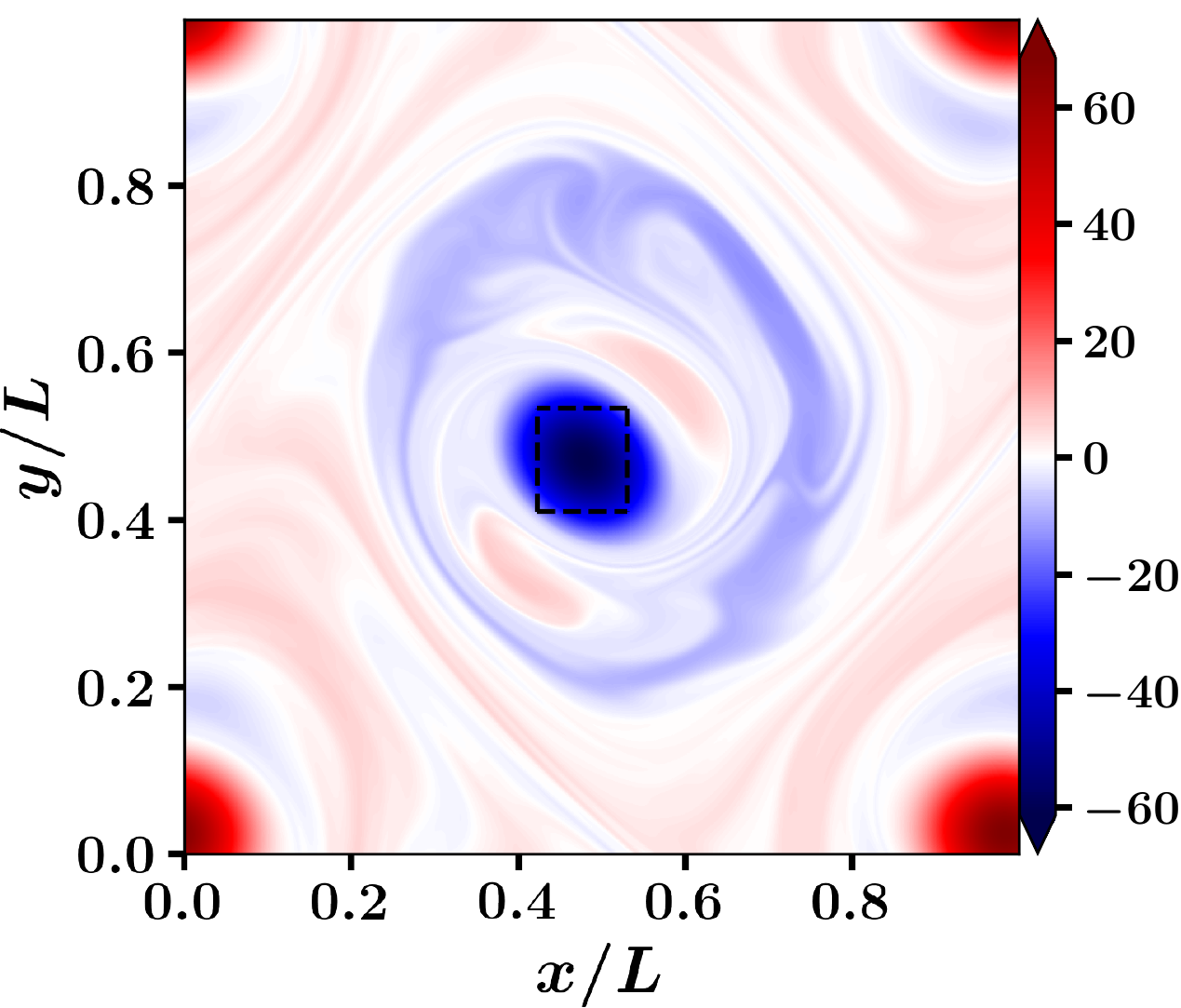}
\includegraphics[scale=0.3]{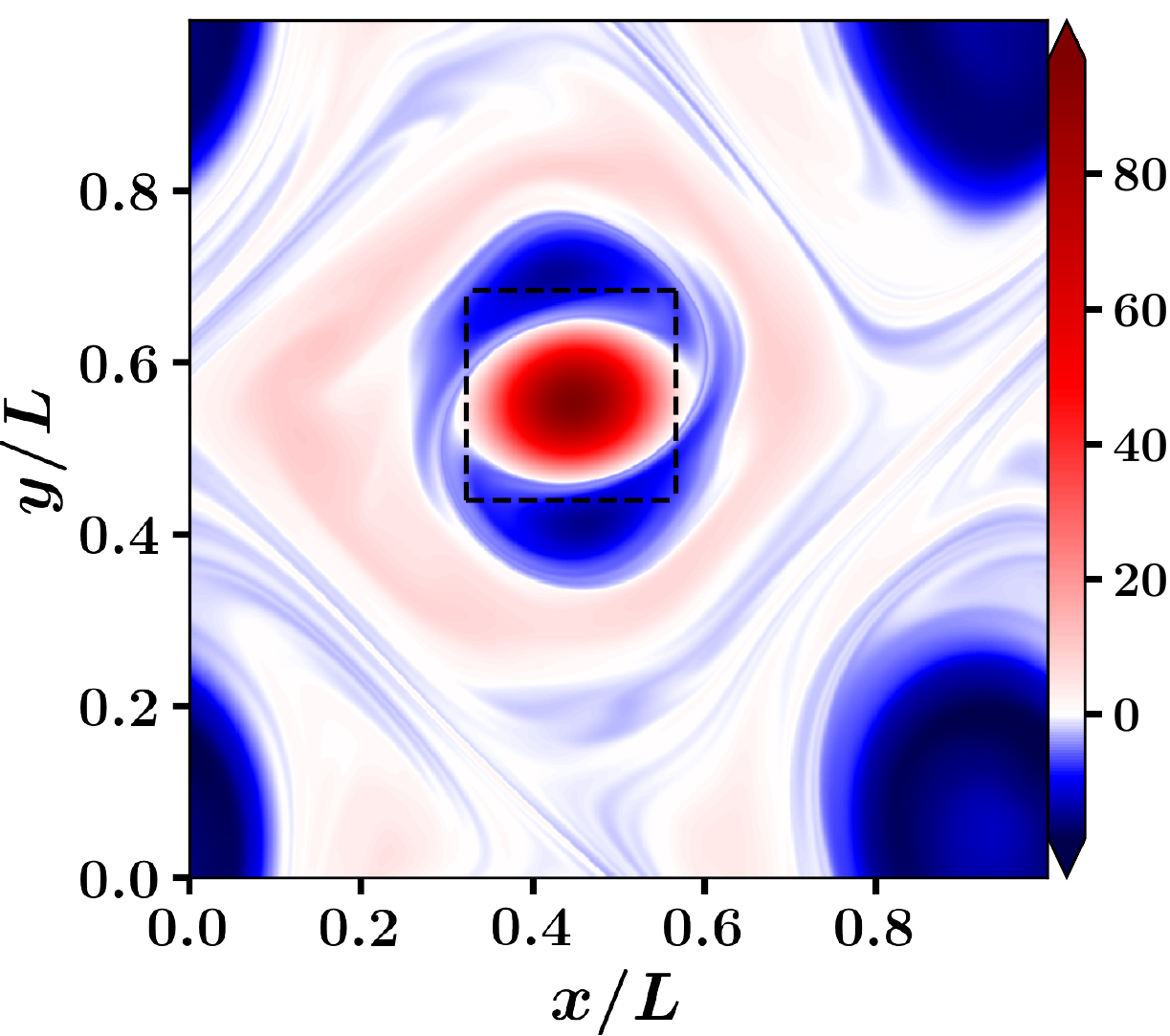}
\includegraphics[scale=0.3]{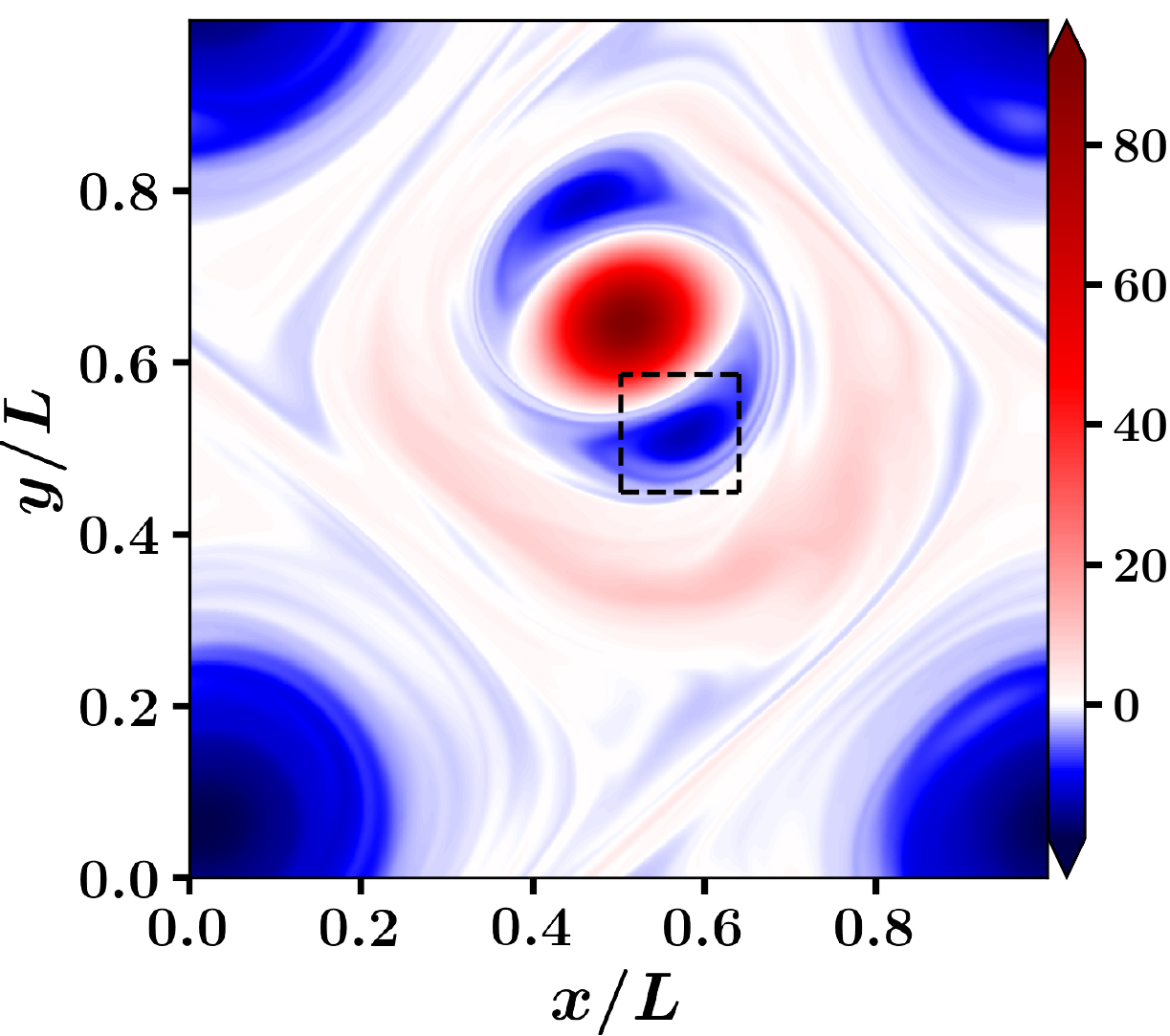}
\end{center}
\begin{center}
\includegraphics[scale=0.3]{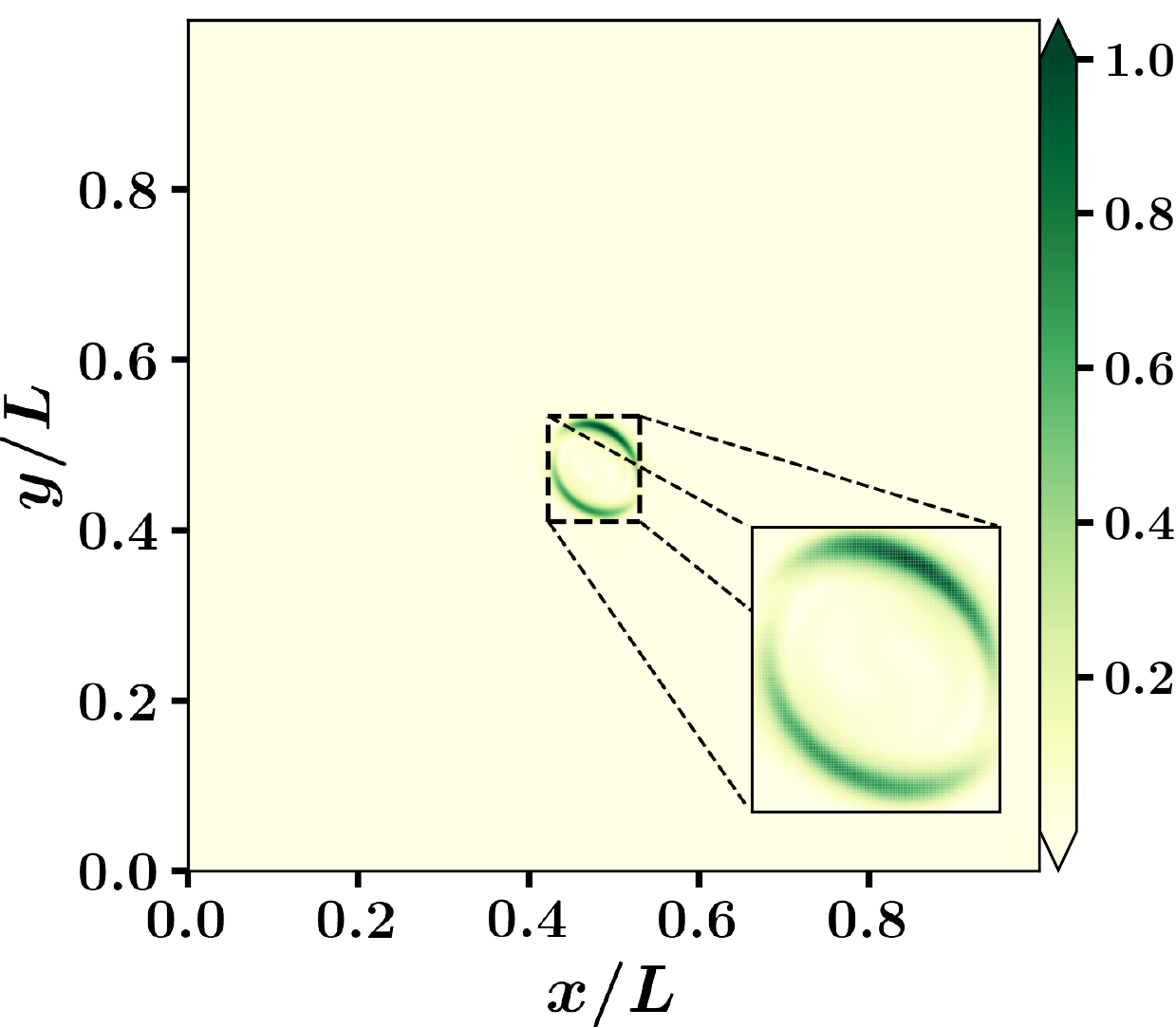}
\includegraphics[scale=0.3]{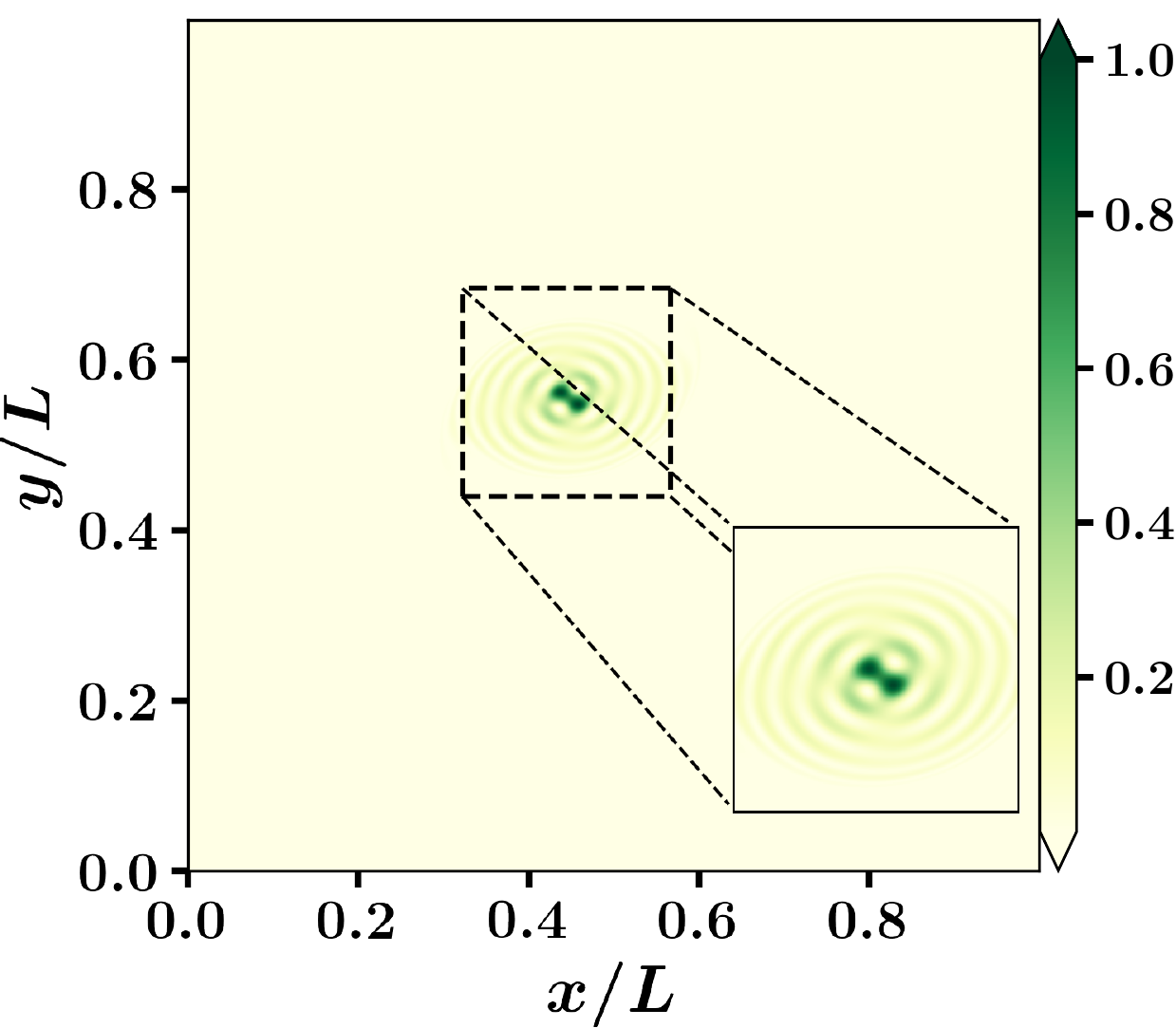}
\includegraphics[scale=0.3]{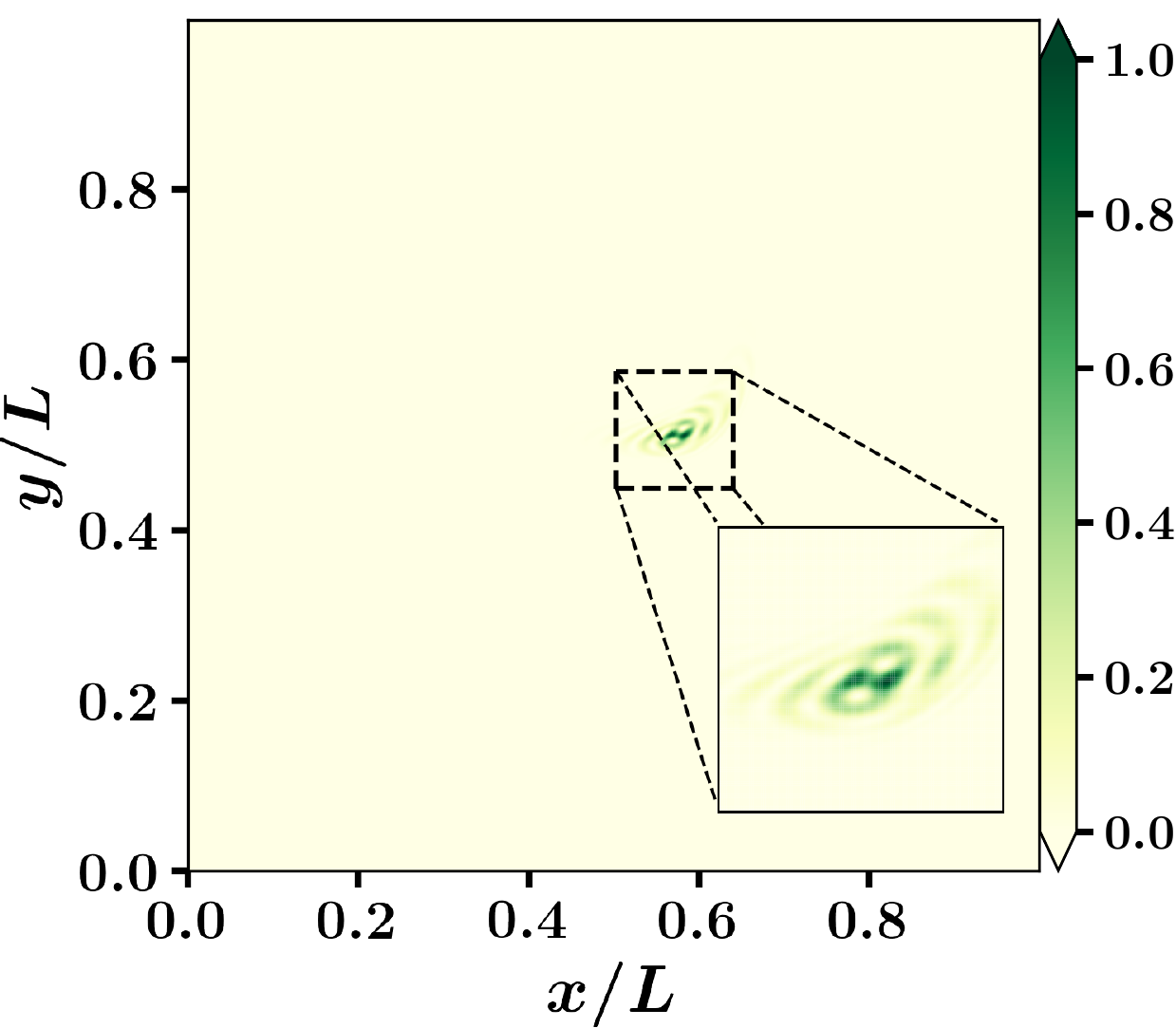}
\end{center}
\caption{\textbf{Unstable mode near threshold.} Top panels: normalised vorticity $\omega_{_{2D}} \frac{L}{U}$ of the 2D turbulent base flow. Bottom panels: corresponding unstable three-dimensional perturbation. Left column: $Re = 9.3\times10^4, q L = 20 \pi, Ro = 2.0\times10^{-2}$. Middle column: $Re = 9.3\times10^4, q L = 6 \pi, Ro = 7.4\times10^{-3}$. Right column: $Re = 9.3\times10^4, q L = 2 \pi, Ro = 2.7\times10^{-3}$.}
\label{fig:uoprofile}
\end{figure*}

\section{Centrifugal instability}

\begin{figure}
\centering{\includegraphics[scale=0.5]{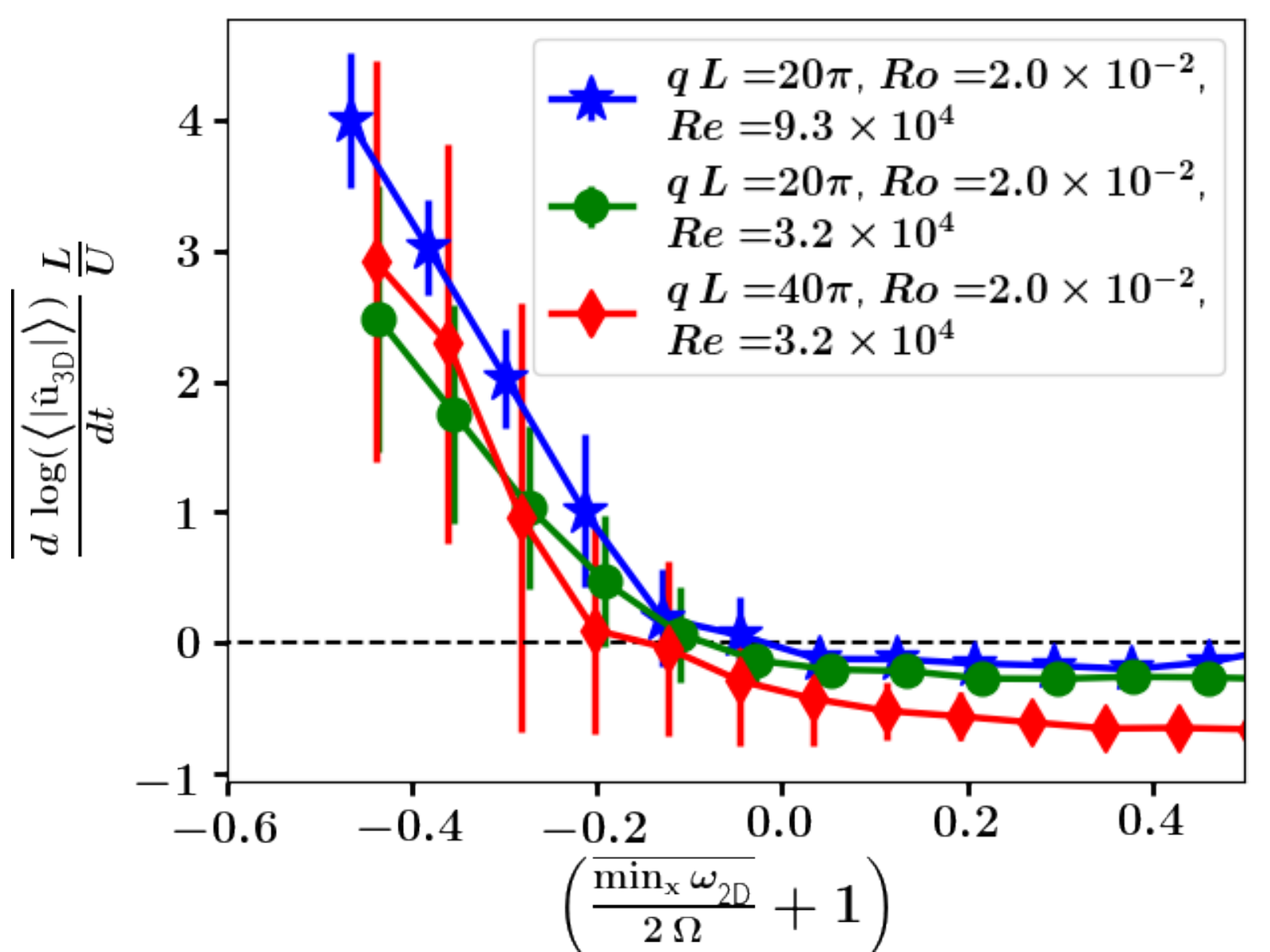}}
\caption{Scatter plot of the smoothed growth rate versus the Rayleigh-like parameter ${\cal R}(t)=\overline{\min_{\bf x} (\omega_\text{2D})}/2\Omega + 1$. The symbols correspond to the mean value in each horizontal bin and the vertical bars show the standard deviation. The strong correlation between the two quantities is characteristic of a centrifugal instability developing inside anticyclones when ${\cal R} \left( t \right) \lesssim 0$.
\label{fig:growthomegamax}}
\end{figure}

For the shallowest fluid domain that we have investigated, $qL=20\pi$, the 3D instability exhibits many features of the centrifugal instability. First, we observe that the unstable mode develops only inside anticyclones, an example being provided in Fig.~\ref{fig:uoprofile} (first column). Secondly, the inviscid Rayleigh criterion for the centrifugal instability of an axisymmetric vortex is that the quantity $\phi(r)=2[V(r)/r + \Omega][\omega_\text{2D}(r)+2 \Omega]$ be negative for some value of $r$, where $V(r)$ and $\omega_\text{2D}(r)$ are the azimuthal velocity profile and the vorticity profile, respectively \citep{kloosterziel1991experimental}. For a given radial structure of the anticyclone, this instability criterion yields $\omega(r=0) + {\cal C} \Omega < 0$, where the first term is the (negative) vorticity at the vortex center, and ${\cal C}>0$ is a dimensionless constant that depends on the shape of the anticyclone \citep{sipp1999vortices,sipp2000three}. As a proxy to the Rayleigh criterion, we have computed the quantity ${\cal R}(t)=[\overline{\min_{\bf x} (\omega_\text{2D})}+2 \Omega]/2\Omega$, where the overline denotes a smoothing over one turnover time $L/U$. In Fig.~\ref{fig:growthomegamax}, we represent a scatter plot of the (smoothed) instantaneous growth rate $\overline{ \frac{\mathrm{d}}{\mathrm{d}t} (\log E_{3D} )}$ as a function of ${\cal R}(t)$. The two quantities are strongly correlated, the growth rate being positive whenever Rayleigh's instability criterion is satisfied, i.e., when ${\cal R}(t) \lesssim 0$. This is another indication that the instability arising in the present system is of centrifugal type. Finally, rigorous upper bound theory indicates that the time average of the minimum 2D vorticity scales with the large-scale turnover time, $| \left< \min_{\bf x} (\omega_\text{2D}) \right>_t | \lesssim U/L$ up to logarithmic terms (see Appendix A. 3 in \cite{gallet2015exact}). This is an indication that $\overline{\min_{\bf x} (\omega_\text{2D})}$ scales as $U/L$, so that the generalised Rayleigh criterion yields $Ro \leq \text{const.}$: at low viscosity (large Reynolds number), we expect the centrifugal instability to arise above a $Re$-independent threshold Rossby number. The shallow-layer $qL=20\pi$ data in Fig.~\ref{fig:RocvsRe} indeed asymptotes to a $Re$-independent threshold Rossby number at large $Re$. A similar correlation between the Rayleigh criterion ${\cal R}(t)$ and the growth rate was observed for all the filled symbols in Fig.~\ref{fig:RocvsRe} with $Ro\lesssim0.1$, which indicates that the corresponding instability is of centrifugal type. However, for large Reynolds number and lower values of the vertical wavenumber, $qL=6\pi$ or $qL=2\pi$ (i.e., for deeper fluid layers), we observed no clear correlations between ${\cal R}(t)$ and the growth rate: these are the open symbols in Fig.~\ref{fig:RocvsRe}. These data points also depart from the $Ro_c=\text{const.}$ asymptote of the shallow layer $qL=20\pi$, which is additional evidence that they do not correspond to a centrifugal instability. 

\section{Three-dimensionality in deeper fluid layers}

For lower values of $qL$ (deeper fluid layers) and large Reynolds number, the threshold to three-dimensionality departs from the $Ro=\text{const.}$ centrifugal asymptote, see Fig.~\ref{fig:RocvsRe}. The deeper the layer, the lower the value $qL$ of the gravest vertical mode, and the sooner the threshold departs from the asymptote as $Re$ increases. We conjecture that the $qL=20\pi$ threshold would probably also depart from the centrifugal asymptote if we could investigate even higher Reynolds numbers. There is thus another instability at play in deep domains and large Reynolds number. To characterise the corresponding unstable modes, in the second and third columns of Fig.~\ref{fig:uoprofile} we provide snapshots of the 2D base flow and 3D perturbation, during a phase of rapid growth of the latter. A first difference with the centrifugal instability is that the unstable mode can develop inside the elliptical core of both cyclones (Fig. \ref{fig:uoprofile}, second column) and anticyclones (Fig. \ref{fig:uoprofile}, third column). This points towards the elliptical instability as a potential candidate for this large-$Re$ instability. However, at least two arguments seem to challenge this interpretation: first, we played the -- arguably artificial -- game of replacing the 2D turbulent base-flow by a steady flow that corresponds to either the top-middle snapshot of Fig. \ref{fig:uoprofile}, or the top-right one (i.e., we freeze the 2D base flow). We probed the stability of the frozen flow over a range of Rossby numbers that extends up to twice the threshold Rossby number of instability of the time-dependent flow. Surprisingly, we observed that such artificial steady flows do not lead to an instability over that range of Rossby numbers. The unavoidable conclusion is that the time-dependence of the base flow plays a central role in the instability mechanism, which thus differs from the simple elliptical instability of a steady vortex. Second, theoretical predictions based on weak-ellipticity expansions lead to a threshold for instability of the form $Re \sim Ro^{-2}$ \citep{le1999short,le2000three}, while our data points indicate a scaling-law closer to $Re \sim Ro^{-1}$ for the 2D-3D threshold at high $Re$ in deep domains, albeit in a moderate parameter range (see Fig.\ref{fig:RocvsRe} for $qL=6\pi$ and $qL=2\pi$).

\begin{figure}
\centering{\includegraphics[scale=0.5]{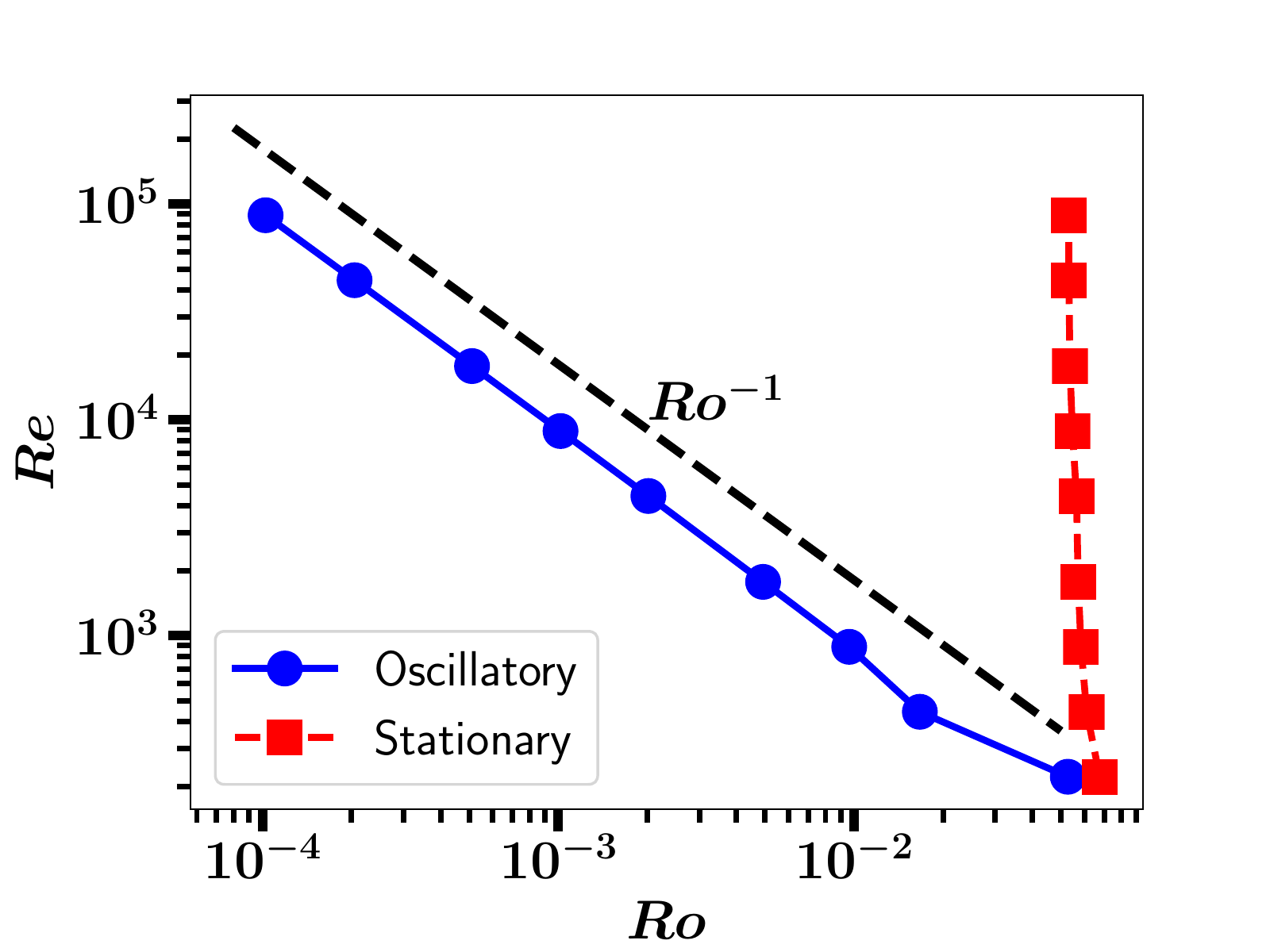}}
\caption{Threshold of instability to 3D perturbations in the $(Ro,Re)$ plane for an oscillatory Kolmogorov flow (blue circles) and a steady Kolmogorov flow (red squares). While the steady Kolmogorov flow has a $Ro \simeq \text{const.}$ instability threshold, the oscillatory Kolmogorov flow is much more unstable, with a threshold Reynolds number that scales as $Ro^{-1}$ (dashed line).\label{fig:Kolmosc}}
\end{figure}

Because the high-$Re$ deep-domain instability arises at low Rossby number, we can address it through a standard low-$Ro$ inertial-wave expansion of the 3D perturbation. In this framework, the 2D base flow and the 3D perturbation are decomposed onto a helical basis of inertial waves, with slowly varying amplitudes. These wave amplitudes evolve primarily through resonant triadic interactions. However, these interactions do not transfer energy between the 3D waves and the 2D base-flow, and therefore the 3D instability cannot arise from resonant triadic interactions \citep{greenspan1968theory,smith1999transfer,Chen2005}. In some sense, this is the reason behind the existence of exact two-dimensionalization: these instabilities, which would lead to a constant threshold Reynolds number $Re_c$ (set by the balance between an inviscid growth rate $\gamma \sim U/L$ and the viscous damping rate $\nu/L^2$) cannot arise over a purely 2D base flow. Instead, one needs to go to the next order in Rossby number to uncover mechanisms that do transfer energy between the 2D flow and the 3D perturbations. At this order, quasi-resonant triads and resonant quartets of inertial waves were recently highlighted as important instability mechanisms to induce 2D motion from an inertial-wave base-flow \citep{smith1999transfer,kerswell1999secondary,leReun2019,leReunthese,Brunet2020,Reun2020}. We argue that the opposite mechanisms can arise in the present system: 3D inertial waves can arise spontaneously over a 2D base-flow, either through resonant quartets of inertial waves, or through the parametric excitation of inertial waves by the time-dependent 2D flow. Both mechanisms would lead to an inviscid growth rate $\gamma \sim Ro \, U/L$: at threshold, the latter balances the viscous damping rate $\nu/L^2$, which leads to a threshold $Re \sim Ro^{-1}$ in parameter space. The parametric excitation of inertial waves by the time-dependent 2D flow is a particularly appealing mechanism, as it provides an explanation to both the $Re \sim Ro^{-1}$ scaling behaviour and the key role of the time dependence of the base flow. A simple illustration of this mechanism is provided by the oscillatory Kolmogorov flow: instead of the intricate 2D turbulent base flow, consider the simpler flow $\ubd= 2 U \sin(4\pi y/L) \cos(4\pi U t/L) {\bf e}_x$ as a model of large-scale flow structures of typical velocity $U$ evolving with the eddy turnover time $L/U$. We determine the threshold for the growth of 3D perturbations around this 2D base flow using a numerical code based on Floquet theory in time. As shown in Fig.~\ref{fig:Kolmosc} for $qL=2\pi$, the threshold of the corresponding parametric instability scales as $Re \sim Ro^{-1}$ (blue circles). By contrast, the steady version of this Kolmogorov flow,  $\ubd= \sqrt{2} U \sin(4\pi y/L) {\bf e}_x$, yields an instability threshold $Ro \simeq \text{const.}$ (red squares), reminiscent of the centrifugal asymptote in figure \ref{fig:RocvsRe}. Indeed, for such a parallel flow the streamlines have an infinite radius of curvature, and the $\phi$ centrifugal-instability criterion above reduces to the existence of a point in the flow where $\omega_\text{2D}\leq - 2 \Omega$. This criterion yields a threshold Rossby number $Ro_c=1/(4\pi \sqrt{2})\simeq 0.0563$ for the steady Kolmogorov flow, in excellent agreement with the numerical value, see Fig.~\ref{fig:Kolmosc}.

The stark contrast between the instability thresholds of the steady and oscillatory Kolmogorov flows in Fig.~\ref{fig:Kolmosc} illustrates once again that the time-dependence of the 2D base flow is a key ingredient for instability to 3D perturbations at high Reynolds number and low Rossby number.

\section{Discussion}

We have investigated the onset of three-dimensionality in rapidly rotating turbulent flows using the capabilities of modern GPUs. Interestingly, the threshold between exactly 2D and partially 3D flows crosses the region of parameter space that is accessible to laboratory experiments and fully 3D DNS. In this region of parameter space, we have provided evidence that three-dimensionality arises through the centrifugal destabilisation of anticyclones. The corresponding threshold Rossby number depends only weakly on the Reynolds number for large enough $Re$. However, our approach also allowed us to reach a region of parameter space that goes beyond the parameter range of state-of-the-art experiments and DNS, up to $Re=10^5$ and $Ro=2 \times 10^{-3}$ simultaneously. For such extreme parameter regimes the threshold between 2D and 3D flows scales as $Re \sim Ro^{-1}$, and the time dependence of the base flow appears as a crucial ingredient of the instability process. We proposed the parametric amplification of inertial waves by the fluctuating 2D turbulent flow as a candidate mechanism for this instability, which we exemplified through the stability analysis of the rapidly rotating oscillatory Kolmogorov flow. Whether and how this instability is connected to standard forms of the elliptical instability remains to be investigated. The scaling $Re \sim Ro^{-1}$ for the instability threshold is yet another confirmation that the limits $Re \to \infty$ and $Ro \to 0$ do not commute: the end state of rapidly rotating turbulence depends very much on the distinguished limit considered when sending $Re$ to infinity and $Ro$ to zero \citep{alexakis2015rotating,gallet2015exact,leReunthese}.

While this study is motivated primarily by theoretical fluid dynamics, some connection can be made to the dynamics of natural flows. Very much like rotating turbulence can be decomposed into a 2D slow manifold coexisting with 3D waves, rotating stratified oceanic and atmospheric flows can be decomposed into balanced motion -- the quasi-geostrophic slow manifold -- and waves. In both cases the slow manifold leads to an inverse energy cascade, with energy condensing into domain-scale structures in the absence of large-scale damping, and in both cases the waves can induce a `wave-turbulent' forward energy cascade. Because such condensation of kinetic energy at the basin scale is absent from oceanic data, it has been hypothesised that part of the balanced energy may be transferred to wave-like motion and cascaded to small scales, a phenomenon coined `loss of balance' \citep{Vanneste2013,SLOB}. In the simpler context of rapidly rotating unstratified turbulence, the slow manifold consists of 2D flows, and loss of balance corresponds to the emergence of 3D waves. Our study thus highlights basic instability mechanisms leading to  spontaneous loss of balance in a rapidly rotating unstratified turbulent flow. It provides the region of parameter space where 3D structures develop, shedding light on the possible emergence of a forward energy cascade as the Rossby number increases.

We stress the fact that the present linear stability analysis only provides sufficient conditions for the emergence of 3D structures. For Rossby numbers lower than $Ro_c$, we cannot rule out the emergence of three-dimensionality through finite-amplitude instabilities (FAIs). In other words, below $Ro_c(Re)$ we know that the system possesses a 2D-flow attractor, but this attractor may coexist in phase space with a fully-3D-flow attractor (see e.g. \citet{yokoyama2017hysteretic} for the coexistence of a quasi-2D attractor and a strongly 3D one). The statistically steady state realised by the system would then depend on the initial condition. As shown in Gallet (2015), however, there exists a value $Ro_\text{abs}(Re)$ of the Rossby number below which such FAIs are ruled out: for $Ro \leq Ro_\text{abs}(Re)$, the 2D flow is absolutely stable to 3D perturbations and the system ends up in the 2D-flow attractor regardless of the initial condition. If $Ro_\text{abs}(Re)<Ro_c(Re)$, then FAIs can arise for $Ro\in[Ro_\text{abs}(Re), Ro_c(Re)]$, depending on the initial condition. If $Ro_\text{abs}(Re)=Ro_c(Re)$, FAIs are ruled out. Again, the determination of $Ro_\text{abs}(Re)$ through fully 3D DNS remains prohibitively expensive at large $Re$ and low $Ro$. Instead, the existence of FAIs, as well as whether and when a forward energy cascade develops, could be investigated through nonlinear extensions of this work: one could design a weakly nonlinear model by keeping only the first unstable vertically dependent mode and its feedback onto the 2D base flow, in the spirit of \cite{benavides2017critical} and \cite{KannaPRL}. For physical systems that are amenable to 3D direct numerical simulation in extreme parameter regimes, models of this kind have predictive skills when compared to 3D DNS, at least at a qualitative level \citep{van2019condensates}. For the problem at stake, such a model may capture the emergence of a forward energy cascade as $Ro$ increases \citep{alexakis2018cascades}, thus providing direct information on the region of parameter space where rapidly rotating turbulence displays an `anomalous' or fully turbulent energy dissipation rate, independent of molecular viscosity.

\acknowledgements{We thank S. Le Diz\`es and A. Alexakis for insightful discussions. This research is supported by the European Research Council (ERC) under grant agreement FLAVE 757239.} 
\\
Declaration of Interests. The authors report no conflicts of interest.

\appendix

\section{Numerical mehods \label{appendixDNS}}

The numerical simulations are performed using standard pseudo-spectral methods, with de-aliasing using the $2/3$-rd rule. The fields are decomposed into a Fourier-Fourier basis in spectral space and discretised on a $(N_x, N_y)$ grid in the $x$ and $y$ directions of the physical space. Time-stepping is performed using a standard four-step third-order Runge-Kutta  scheme. The adaptive time step $dt$ satisfies a CFL condition constructed from both advective and rotation time scales. The spatial resolutions and the average timestep are given in Table \ref{tbl:numsims}. The simulations are run for a number of timesteps of the order of $Re L /(U dt)$.
\begin{table}
\begin{center}
\begin{tabular}{  p{3.5cm} p{3.5cm} p{2.5cm} p{3cm} }
 $Re$ & $q\, L$  & Resolution $(N_x, N_y)$ & $dt \, \frac{U}{L}$\\ \hline
 $\leq 5 \times 10^2$ & $[2 \, \pi, 6 \, \pi, 20 \, \pi]$ & $256 \times 256$ & $[3 \times 10^{-4}, 5\times 10^{-4}]$ \\  
 $[1,2 \times 10^3,2.5 \times 10^3]$ & $[2 \, \pi, 6 \, \pi, 20 \, \pi]$ & $384 \times 384$ & $2 \times 10^{-4}$  \\ 
 $1.2 \, \times 10^4$ &  $[2 \, \pi, 6 \, \pi, 20 \, \pi]$ & $512 \times 512$ & $10^{-4}$  \\  
 $3.2 \, \times 10^4$ & $[2 \, \pi, 6 \, \pi, 20 \, \pi]$ & $1024 \times 1024$ & $7 \times 10^{-5}$  \\ 
 $9.3 \, \times 10^4$ & $[2 \, \pi, 6 \, \pi, 20 \, \pi]$ & $1536 \times 1536$ & $4 \times 10^{-5}$  \\  \hline
\end{tabular}
\caption{Resolution $(N_x, N_y)$ and average (adaptive) time step $dt \, \frac{U}{L}$ for the values of the Reynolds number considered in this study. The simulations are run for a large number of timesteps, of order $Re L /(U dt)$.}
\label{tbl:numsims}
\end{center}
\end{table}

Simulations of the 2D Navier-Stokes equations (\ref{NS2D}) are performed until the flow reaches a statistically steady state, which constitutes the base flow of the present linear stability analysis. The Reynolds and Rossby numbers are built using the rms velocity $U$ of this 2D base flow. The control parameter of the 2D Navier-Stokes equation (\ref{NS2D}) is the Grashof number $Gr=f_0^{1/2} L^{3/2} /\nu$, while the rms velocity $U$ is an emergent quantity. For low Grashof number, $U$ is given by the laminar balance between the viscous and forcing terms, $U/\sqrt{f_0 L} \sim Gr$. For larger Grashof number, $Gr \gtrsim 500$, the rms velocity obeys the approximate scaling-law $U/\sqrt{f_0 L} \simeq 6.0 \times 10^{-3} \,Gr^{1/2}$ with 20$\%$ accuracy.



\bibliographystyle{jfm}
\bibliography{apssamp}

\end{document}